# Electronic nonequilibrium effect in ultrafast-laser-irradiated solids

Nikita Medvedev[1,2,*]

*1) Institute of Physics, Czech Academy of Sciences, Na Slovance 1999/2, 182 21 Prague 8, Czech Republic*
*2) Institute of Plasma Physics, Czech Academy of Sciences, Za Slovankou 3, 182 00 Prague 8, Czech Republic*

## Abstract

This paper describes the effects of electronic nonequilibrium in a simulation of ultrafast laser irradiation of materials. The simulation scheme based on tight-binding molecular dynamics, in which the electronic populations are traced with a combined Monte Carlo and Boltzmann equation, enables the modeling of nonequilibrium, nonthermal, and nonadiabatic (electron-phonon coupling) effects simultaneously. The electron-electron thermalization is described within the relaxation-time approximation, which automatically restores various known limits such as instantaneous thermalization (the thermalization time $\tau_{e-e} \to 0$) and Born-Oppenheimer (BO) approximation ($\tau_{e-e} \to \infty$). The results of the simulation suggest that the non-equilibrium state of the electronic system slows down electron-phonon coupling with respect to the electronic equilibrium case in all studied materials: metals, semiconductors, and insulators. In semiconductors and insulators, it also alters the damage threshold of ultrafast nonthermal phase transitions induced by modification of the interatomic potential due to electronic excitation. It is demonstrated that the models that exclude electron-electron thermalization (using the assumption of $\tau_{e-e} \to \infty$, such as BO or Ehrenfest approximations) may produce qualitatively different results, and a reliable model should include all three effects: electronic nonequilibrium, nonadiabatic electron-ion coupling, and nonthermal evolution of interatomic potential.

## I.    Introduction

Ultrafast laser irradiation of materials plays an important role in both, fundamental and applied sciences [1–5]. Understanding of basic phenomena in the physics of solids, nonequilibrium kinetics, and highly excited states of matter, benefits from experiments accessing the natural time window of the involved processes (femtosecond timescales) such as electron kinetics, electron-ion (electron-phonon) coupling, and atomic response [5–7]. Laser irradiation experiments are vital for materials processing, nano and micro technology, and medical applications such as laser surgery [4,5,8,9]. In turn, the interpretation of experimental results requires detailed theoretical and model descriptions of the processes involved.

When an intense ultrafast laser pulse arrives at a material, a sequence of processes takes place, ultimately leading to material modification and the formation of a new state [1,2,10]. It starts with photoabsorption, which, for photon energies above the bandgap, is performed primarily by electrons of the material with the highest ionization potential among those smaller than the photon energy [11]. The photoelectrons are excited to higher energy states, which drives the electronic ensemble out of equilibrium [11,12]. The electrons then scatter among themselves in the conduction band of material, with electrons from the valence band and deeper shells (for those shells whose ionization potential is below the electron energy), or elastically with the atoms of the target (also known as the electron-phonon scattering). Electron-electron scattering thermalizes the electronic ensemble towards equilibrium Fermi-Dirac distribution, whereas electron-atom scattering exchanges energy between the two systems, heating the lattice. Typical electronic thermalization in metals takes place at femtosecond timescales [1,13], while equilibration between the electronic and atomic temperatures may take up to tens of picoseconds [1,14].

---

* Corresponding author: ORCID: 0000-0003-0491-1090, email: nikita.medvedev@fzu.cz

At the same time, since the electrons form the interatomic potential of the material, the excitation of electrons modifies the potential energy surface [1,10,15]. Atoms, in their former equilibrium positions, now experience new forces, which may drive them into a different material phase – the processes known as nonthermal phase transitions, the most famous example of which is nonthermal melting in covalently bonded solids [16,17]. Furthermore, in the time window when the electronic temperature (or, more generally, parameters of the nonequilibrium electronic distribution) is different from the atomic one, this so-called two-temperature state departs from the equilibrium material phase diagram and may create states inaccessible under equilibrium conditions [18,19].

The variety of processes, taking place after ultrafast irradiation, may create synergy leading to nonlinear behavior [20]. Previously, the nonthermal phase transitions were studied [21–23], in a few cases with nonadiabatic coupling included (i.e., the energy exchange between electrons and atoms induced by atomic motion, such as the electron-phonon coupling) [10,24]. Effects of the electronic nonequilibrium were also analyzed separately [1,13,25,26]. The mutual effect of the electronic nonequilibrium and the electron-phonon coupling has been demonstrated [27,28]. However, a combined effect of the interplay of the electronic nonequilibrium, nonadiabatic coupling, and nonthermal effects, was not studied before.

In this work, the interplay of these effects is investigated using the developed combined model that simultaneously traces the kinetics of electrons, the evolution of the electronic structure, and the response of the atomic system. The model for such a description is presented in section II below. In section III.1, it is demonstrated that electron nonequilibrium slows down electron-phonon coupling, leading to prolonged heating of the atomic system in metals. A validation of the model against an experiment is also presented there. In semiconductors, discussed in Section III.2, electronic nonequilibrium enhances nonthermal effects, altering their damage threshold. Section III.3 describes the response of insulators to irradiation, demonstrating the importance of the inter-band interaction of electrons, which leads to thermalization between the valence and the conduction band electrons.

## II.    Model

To model the response of materials to laser irradiation, the XTANT-3 hybrid code [29] is modified to account for electronic nonequilibrium among the low-energy electrons populating the valence and the conduction bands [10]. The code combines a few different models to simultaneously trace the essential effects in both, the electronic and the atomic systems of the target. (i) The electrons with energy above a certain chosen energy cutoff are modeled with the Monte Carlo (MC) simulation. (ii) The fractional populations of low-energy electrons in the valence and conduction band energy levels are traced with the Boltzmann equation (BE). (iii) The interatomic forces are calculated from the transferable tight-binding (TB) formalism, which also traces the evolution of the electronic energy levels (molecular orbitals). (iv) The motion of atoms is modeled with the help of the molecular dynamics simulation (MD). Below, we will discuss the relevant details of each model, and their interconnection on the fly, enabling to model laser irradiation of matter.

(i) In the current implementation, photoabsorption is modeled in the linear regime only, assuming the photon energy is sufficiently high or, for low photon energies, the laser intensity is sufficiently low to neglect nonlinear effects – namely, simultaneous multiphoton absorption (or absorption via virtual levels) is not included [30]. However, a sequential multiphoton absorption is possible in this model, when the same electron may absorb a few photons one by one being sequentially excited to high-energy states in the conduction band. The photoabsorption is modeled with the Monte Carlo simulation, which uses photoabsorption cross sections from the EPICS2017 database for deep atomic shells (possible for X-ray irradiation) [31], whereas for the valence or conduction bands photoabsorption may be included *via* complex dielectric function [32].

Electrons above a chosen cutoff energy (typically, 10 eV counted from the bottom of the conduction band) are also treated with the MC module as free electrons [33]. Such a separation of electrons into the low-energy fraction and the high-energy fraction is necessary within the model for several reasons. Since TB energy levels are calculated as the linear combination of atomic orbitals, they are typically limited to energies of about 10 eV, and higher energy electrons cannot be reliably traced. On the other hand, modeling high-energy electrons as free particles is a standard approximation in transport MC methods and is significantly easier and faster than more advanced methods such as time-dependent ab initio simulations [34].

The event-by-event MC scheme relies on the asymptotic trajectory method. The impact ionization and (quasi-)elastic scattering processes are included. The impact ionization is modeled with the binary-encounter Bethe (BEB) cross-section [35], or, alternatively, with the linear response theory [36]. The elastic scattering is modeled with the screened Rutherford (Mott) cross-section with the modified Molier screening parameter [37].

Upon impact ionization, a new electron may be excited above the cutoff energy. In such a collision, the choice of an electron being excited is made taking into account the transient distribution function of low-energy electrons, evolving in time according to the Boltzmann equation described below, in item (ii). An excited secondary electron is traced in MC simulation in the same way as the primary photoelectron. Each scattering of an electron on the valence- and conduction-band electrons provides the low-energy electron fraction with energy and may excite new electrons. This changes the number of particles and energy content, modifying the low-energy electron distribution function at each time step of the simulation. Also, when a high-energy electron loses energy below the cutoff, it disappears from the MC module and joins the low-energy electrons populating the valence and the conduction bands traced with the BE, as will be described below in item (ii).

In an elastic scattering event, no ionization is produced but the energy is transferred to the atomic system as an increase in the kinetic energy of atoms. The energy transfer is traced from all electrons scattering within the given MD timestep and averaged over the MC iterations.

Core shells, created by X-rays or impact ionization by fast electrons, decay *via* the Auger channel with the characteristic times from the EPICS2017 database [31]. In such a process, two new holes are created, which, if in core shells, will undergo their own Auger decays. High-energy Auger electrons, emitted in decays, are traced in the MC module.

(ii) To describe the dynamics of (low-energy) electrons and atoms, one may start with the Ehrenfest approximation (and modify it as necessary) [38]. It is based on the mean-field approach, allowing a reduction of the problem to quantum electrons and classical atoms. Electrons are then described with the equation for the evolution of the density matrix, whereas expectation values are tracked for atoms, which reduces to the classical (Newton's) equations with the potential defined by the transient state of the electronic system. It is further assumed that the diagonal elements of the density matrix – the electron populations or distribution function, $f_e(\varepsilon_i, t)$, – may be traced with a semiclassical Boltzmann equation. This assumption also requires statistical averaging over the electronic ensemble, which makes the electron populations fractional. The off-diagonal elements are then only represented as the scattering integrals, defining the evolution of the electronic distribution:

$$\frac{df_e(\varepsilon_i, t)}{dt} = I_{e-e} + I_{e-a} + I_{MC}. \tag{1}$$

The distribution function describes fractional electronic populations in states associated with discrete energy levels $\varepsilon_i = \langle i | H_{TB} | i \rangle$, which are the eigenfunctions of the TB Hamiltonian (described below) at the current MD timestep. Here $I_{e-e}$ is the electron-electron collision integral responsible for electronic thermalization; $I_{e-a}$ is the electron-atom collision integral defining the nonadiabatic energy exchange between electrons and atoms induced by atomic motion (e.g., electron-phonon coupling); $I_{MC}$ is the source term describing the change of the distribution function induced by the processes accounted for in the MC module (photoabsorption, Auger-decays involving valence/conduction band, high-energy electrons scattering and influx). All these collisional and source terms are updated at each time step of the simulation.

The standard Ehrenfest dynamics, being a mean-field approximation, does not by itself account for electronic thermalization. Thus, in Eq.(1), the term $I_{e-e}$ is added *ad hoc*, knowing that the Boltzmann equation must lead to thermalization.

For the electron-atom scattering, $I_{e-a}$, the model developed in Ref. [14] is employed. It is a semiclassical Boltzmann collision integral, in which the probability of electronic transition between energy levels induced by atomic displacement at each MD timestep is calculated from the overlap of the electronic wavefunctions, obtained within the TB method (described below, in item (iii)). The classical Maxwellian distribution is used for atoms with the kinetic temperature calculated directly from the MD velocities, whereas for electrons, Pauli blocking factors are included [14].

For the electron-electron scattering, $I_{e-e}$, the relaxation time approximation is used in the current implementation [39]:

$$I_{e-e} = -\frac{f_e(\varepsilon_i, t) - f_{eq}(\varepsilon_i, \mu, T_e, t)}{\tau_{e-e}}. \tag{2}$$

Here $\tau_{e-e}$ is the characteristic electron-electron relaxation time; $f_{eq}(\varepsilon_i, \mu, T_e, t)$ is the equivalent equilibrium Fermi-Dirac distribution with the same total number of (low-energy) electrons ($n_e$) and energy content ($E_e$) as in the transient nonequilibrium distribution:

$$\begin{cases} n_e = \sum f_e(\varepsilon_i, t) = \sum f_{eq}(\varepsilon_i, \mu, T_e, t) \\ E_e = \sum \varepsilon_i f_e(\varepsilon_i, t) = \sum \varepsilon_i f_{eq}(\varepsilon_i, \mu, T_e, t) \end{cases}. \tag{3}$$

Eqs. (3) define the equivalent electronic temperature (also called the kinetic temperature, $T_e$ [40]) and the equivalent chemical potential ($\mu$). Within this ansatz, the total number of low-energy electrons and the total energy (in electrons and atoms) are conserved within an MD timestep (changes in the number and energy of electrons may only occur *via* the $I_{MC}$ term).

Eqs. (1,2) naturally unify various widely used approaches to quantum-classical dynamics. It recovers the limiting cases of the Born-Oppenheimer (BO) molecular dynamics (in the limit of infinite electronic thermalization time, $\tau_{e-e} \to \infty$, and no nonadiabatic electron-atom coupling, $I_{e-a} = 0$) in which the electronic populations are fixed; the Ehrenfest dynamics which includes average electron-atom energy exchange but no electron thermalization ($\tau_{e-e} \to \infty$, $I_{e-a} \neq 0$); instantaneous thermalization in the adiabatic microcanonical ensemble ($\tau_{e-e} = 0$, $I_{e-a} = 0$, used e.g. in Refs.[33,41]); and nonadiabatic dynamics with instantaneous electron thermalization ($\tau_{e-e} = 0$, $I_{e-a} \neq 0$, used in the previous versions of XTANT [10,14]; the same assumptions are used in the two-temperature based models, TTM-MD [42–44]). Let us also note that the BO approximation only assumes the decoupling of the electronic and atomic wave functions, but not necessarily the ground state – excited adiabatic states may also be calculated in many cases, where potential energy surfaces are far apart, thus suppressing electronic transitions between them (far from such situations as conical intersections or avoided crossings) [45]. The ground-state BO molecular dynamics is a separate additional approximation, which can be reproduced within the relaxation-time formalism as the zero-temperature instantaneous thermalization with no coupling ($\tau_{e-e} \to \infty$, $I_{e-a} = 0$, and $T_e = 0$).

The influx or outflux of electrons from the valence and conduction bands (low-energy electrons) *via* such processes as photoabsorption, scattering of high-energy electrons, and Auger decays, are traced in the MC simulation, information about which is delivered to the Boltzmann equation at each timestep. In each scattering act involving the valence and conduction electrons, the energy level involved is sampled and recorded into $I_{MC}$. The levels are sampled according to probability following the Pauli blocking term in the Boltzmann collision integral: $w_i \sim f(\varepsilon_i)(2 - f(\varepsilon_i + \Delta E))$, where $\Delta E$ stands for the energy transferred in the collision under consideration (factor 2 is due to spin degeneracy). All the changes in populations in each energy level within the given timestep are then averaged over the concurrently executed MC iterations.

A single discrete level from which an electron is emitted (in photoabsorption, impact ionization, or Auger decay of a core hole) can be chosen. However, an incoming electron generally comes with energy somewhere in between the discrete energy levels (since the energy grid cannot be chosen arbitrarily but is unambiguously defined as the eigenstates of the transient Hamiltonian). In this case, the influx of the particles and energy is distributed between the two closest levels under the condition of conservation of the number of particles and energy: increasing the total number of low-energy electrons by one out of $N_{MC}$ iterations and bringing energy $\Delta E$ in the scattering event:

$$\begin{cases} \Delta f_e(\varepsilon_i, t) = \Delta n \dfrac{\varepsilon_j - \Delta E}{\varepsilon_j - \varepsilon_i} \\ \Delta f_e(\varepsilon_j, t) = \Delta n \dfrac{\Delta E - \varepsilon_i}{\varepsilon_j - \varepsilon_i}' \end{cases} \tag{4}$$

where for one incoming electron out of $N_{MC}$, the change in the number is $\Delta n = 1/N_{MC}$, and the corresponding energy change is $\Delta E = E_{el}/N_{MC}$ ($E_{el}$ is the energy brought by this electron). In this way, the total number of electrons and energy, in the low- *and* high-energy fractions (MC and BE modules), are conserved. In the case when one of the levels ($i$ or $j=i+1$) is fully occupied, another set of levels can be chosen – Eqs. (4) hold for arbitrary levels $i$ and $j$. The total change of the population on each level is then summed over all scattering acts within the given timestep: $I_{MC} = \sum \Delta f_e(\varepsilon_i, t)$.

(iii) The tight-binding method is used to calculate energy levels and interatomic forces (potential energy surface) with the Slater-Koster approximation [46]. To make the method transferable to multiple atomic structures, the radial part in the hopping integrals is a function depending on the interatomic distance [47]. For elemental metals, we use NRL TB parameterization [48,49], for Si the parameterization from Ref. [50] is used, and for compounds, we employ DFTB matsci-0-3 parameterization [51]. Nonorthogonal TB Hamiltonian is constructed on each MD timestep of the simulation and diagonalized by solving the secular equation [47]. TB wavefunctions (eigenfunctions of the transient Hamiltonian) can be used to calculate the probabilities of the nonadiabatic electron coupling to atomic motion [14]. Unfortunately, the tight binding method does not allow evaluating electron-electron scattering probabilities at the same level of theory, hence the characteristic relaxation time $\tau_{e-e}$ is an external parameter. Note, however, that further approximations can, in principle, be made for calculations of the electron-electron scattering probabilities or cross sections (such as linear response theory), which may be the subject of a future work.

(iv) Molecular dynamics module in XTANT-3 uses Martyna-Tuckerman's $4^{th}$ order algorithm to propagate classical trajectories of atoms typically with the timestep smaller than 1 fs [52]. Over 200 atoms are used in each simulation box with periodic boundary conditions to mimic bulk materials. No energy sinks associated with particles and energy transport out of the simulation box are included, assuming homogeneous material excitation in all presented simulations.

The forces acting on the atoms are calculated as the gradients of the potential energy defined within the TB formalism:

$$V(\{R_{ij}(t)\}, t) = E_{rep}(\{R_{ij}(t)\}) + \sum_i f(\varepsilon_i, t)\varepsilon_i, \qquad (5)$$

where the potential $V$ depends on pairwise distances between all the atoms in the simulation box $\{R_{ij}(t)\}$; $E_{rep}$ is the effective ion-ion TB repulsion term (containing all contributions apart from the electronic band energies, if required in the given TB parameterization). Additionally, the nonadiabatic energy transfer from the Boltzmann collision integral and the energy transfer from high-energy electrons elastic scattering are fed to atoms *via* velocity scaling at each timestep, which ensures the energy conservation in the entire system: all electrons and atoms (microcanonical ensemble).

Using Eq.(1) allows directly tracing an effect of the nonequilibrium electronic distribution function on the interatomic forces in Eq.(5) (as well as on the electron-atom coupling, $I_{e-a}$ [14]), and thus on the dynamics and stability of the material. As mentioned above, various approximations for the relaxation time also enable comparisons of different standard methods (e.g., BO, instantaneous thermalization approximation).

The hybrid approach of XTANT-3 is sufficiently general to apply to many classes of materials. With all the other modules in the code being universal, the major bottleneck of the method is the transferrable tight binding parameters. Parameterizations such as NRL [48] and DFTB [53] allow to describe a wide range of elemental solids and compounds, except for such complex systems as strongly correlated systems, superconductors, or complex biological samples, which require more refined treatment of the electronic structure than the tight binding method [34].

Note that the formalism, at its core, relies on the extended Ehrenfest dynamics, and not on the finite-temperature extension to *ab initio* simulations [54]. In the finite-temperature models, it is assumed that the constant electronic temperature is enforced by the interaction of the electronic ensemble with the bath. In contrast, in our case, even in the limit of the instantaneous electron thermalization ($\tau_{e-e} \rightarrow 0$), no interaction with the bath is assumed; the electronic thermalization is an intrinsic process within the electronic ensemble considered (non-mean-field effect added). The electronic temperature and chemical potential are not variables in this formulation but parameters.

Let us also note that Eq.(2) could be replaced with the electron-electron Boltzmann collision integral, which does not employ any electronic temperature or chemical potential, even as parameters of the equivalent equilibrium distribution [13]; it could also eliminate the need for the additional parameter – the electron thermalization time – but would require some approximations for the electron-electron scattering probability or matrix element, which is beyond the scope of the current work.

## III.  Results

### 1.  Metals

We start with the study of the evolution of the electron distribution function in irradiated metals in different approximations for the electron-electron characteristic relaxation time. For the study, we chose the photon energy of 2 eV (wavelength of 620 nm) for metal irradiation; a soft X-ray irradiation will be also discussed below, and the effects of various photon energies will be compared for the case of semiconductors in section III.2. In all simulations, the temporal profile of the laser pulse is assumed to be Gaussian with 10 fs FWHM centered at 0 fs.

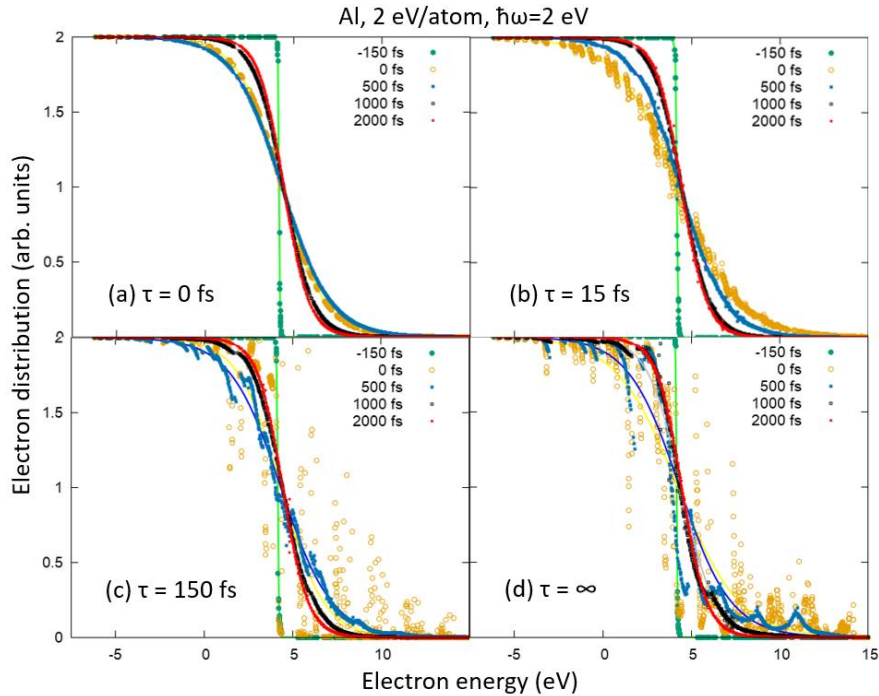

*Figure 1. Electronic distribution function in aluminum irradiated with a pulse of 2 eV/atom, 2 eV photon energy, 10 fs FWHM. Simulations performed with different characteristic electronic relaxation times (here and in all the figures τ stands for $τ_{e-e}$): (a) $τ_{e-e}=0$ fs (instantaneous relaxation); (b) $τ_{e-e}=15$ fs; (c) $τ_{e-e}=150$ fs; (d) $τ_{e-e}→∞$ (no electron-electron relaxation; Ehrenfest-like approximation). Symbols are the transient distribution functions, lines are the equivalent Fermi distributions.*

Figure 1 shows the electronic distribution function in aluminum under irradiation with the optical pulse simulated with various approximations: $τ_{e-e} = 0$ fs (instantaneous electron equilibration), $τ_{e-e} = 15$ fs, $τ_{e-e} = 150$ fs, and $τ_{e-e} → ∞$ (no electron-electron relaxation; Ehrenfest-like approximation). In the instantaneous thermalization limit ($τ_{e-e} = 0$ fs, Figure 1a), the distribution function coincides with the equilibrium Fermi distribution at all times. In simulations with finite thermalization times, the nonequilibrium electronic distribution function transiently forms a staircase-like shape, in agreement with the previous studies [13]. When fast electronic thermalization is assumed ($τ_{e-e} = 15$ fs, Figure 1b), the distribution function quickly turns into equilibrium Fermi function. With longer characteristic electron-electron thermalization times ($τ_{e-e} = 150$ fs, Figure 1c), the deviations from the Fermi functions are stronger and the equilibration takes proportionally longer. At the infinite thermalization, even more pronounced peaks form in the transient distribution, resulting from the convolution of the initial and final states participating in the photoabsorption, calculated at the gamma-point band structure.

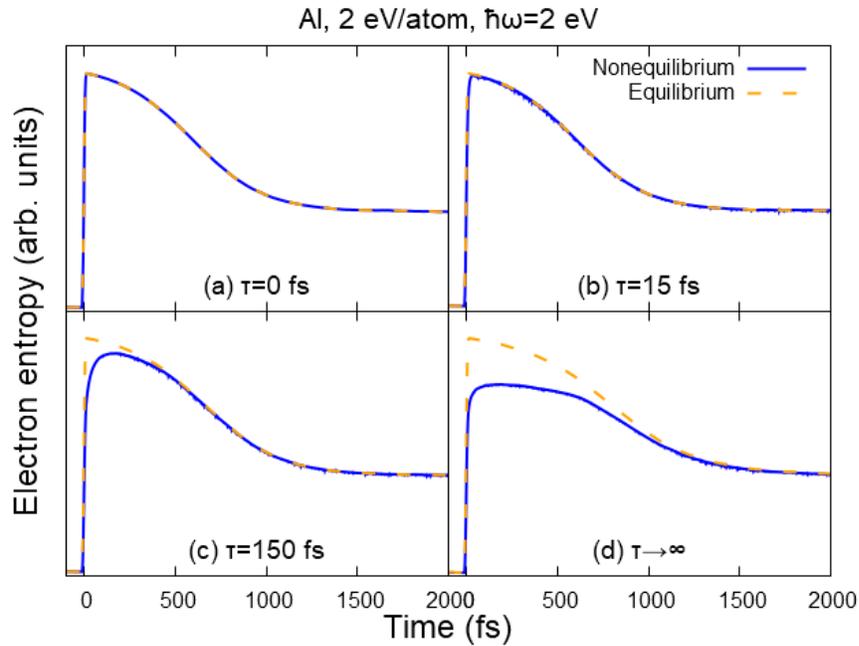

Al, 2 eV/atom, ℏω=2 eV

*Figure 2. Evolution of the electronic entropy in aluminum irradiated with a pulse of 2 eV/atom, 2 eV photon energy, 10 fs FWHM. Simulations performed with different characteristic electronic relaxation times: (a) $τ_{e-e}=0$ fs (instantaneous relaxation); (b) $τ_{e-e}=15$ fs; (c) $τ_{e-e}=150$ fs; (d) $τ_{e-e}→∞$ (no electron-electron relaxation; Ehrenfest-like approximation). Solid lines are the entropies corresponding to the transient distribution functions, dashed lines are the entropy associated with the equivalent Fermi distributions (maximal entropy).*

The most surprising observation is that the Ehrenfest-like dynamics (with no electron-electron scattering, $τ_{e-e} → ∞$, Figure 1d) also leads to the eventual thermalization of electrons via electron-phonon scattering, when the kinetic temperatures of electrons (the equivalent temperature of the equilibrium distribution function according to Eq.(3)) and ions equilibrate. This effect is quantified in Figure 2, where the electronic entropy is shown ( $S_e = -k_B 2 \sum [(f_e/2) \ln(f_e/2) + (1 - f_e/2) \ln(1 - f_e/2)]$, with $k_B$ being the Boltzmann constant, and factors of 2 due to the normalization of the distribution function according to the spin degeneracy). For short electron-electron thermalization times, the entropy quickly reaches the maximal value (Figure 2a and b), which corresponds to the equilibrium distribution function. In all cases, the maximal entropy decreases with time due to the cooling down of the electronic ensemble *via* electron-phonon coupling. In the Ehrenfest-like approximation ($\tau_{e-e} \rightarrow \infty$), the entropy coincides with the maximal one at the time of ~1.5 ps, indicating electronic thermalization.

It is important to keep in mind that this is only an effect of one channel of electron interaction, and, despite the apparent thermalization, the actual channel responsible for the equilibration of electrons is missing in this approximation. Including electron-electron interaction leads to much faster electronic thermalization (compare, e.g., Figure 2b vs. Figure 2d).

The nonequilibrium electron distribution affects the electron-phonon coupling, and thereby the atomic temperature evolution in irradiated aluminum, see Figure 3. The electronic temperature meets the atomic one after equilibration at the times of ~1.5 ps. The drop of the electronic temperature from the peak is larger than the corresponding rise of the atomic temperature, defined by their respective heat capacities.

The thermalized electrons ($\tau_{e-e} = 0$ fs) couple to ions/phonons most efficiently, resulting in the fastest increase of the atomic temperature, as was also noted in previous works [13]. Strongly nonequilibrium electronic ensemble ($\tau_{e-e} \rightarrow \infty$) produces the smallest electron-phonon coupling and, correspondingly, the slowest atomic heating. The coupling parameter depends on both, the electronic and atomic temperature (among other parameters such as material structure and density), and thus increases with time proportionally to the atomic temperature, see Figure 3b [14]. The kinetic (equivalent) temperatures in the case of nonequilibrium distributions are used to define the coupling parameter in Figure 3b; also note that it is only defined for $T_e \neq T_a$, thus the timescale in Figure 3b is limited to before equilibration of the electronic and atomic temperatures.

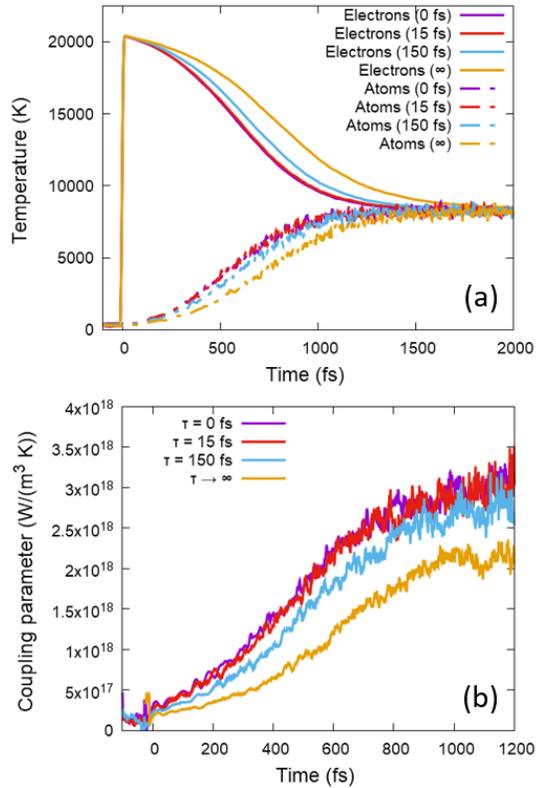

*Figure 3. (a) Electronic and atomic temperatures, and (b) electron-phonon coupling parameter in aluminum irradiated with a pulse of 2 eV/atom, 2 eV photon energy, 10 fs FWHM. Simulations performed with different characteristic electronic relaxation times: $\tau_{e-e}$=0 fs, 15 fs, 150 fs, and ∞ (no electron-electron relaxation; Ehrenfest-like approximation). Solid lines are the electronic kinetic (equivalent) temperatures, and dashed lines are the atomic kinetic temperatures.*

The difference in atomic temperatures, however, is not large among all the simulations, and by the time of ~1500 fs, the atomic temperature reaches the plateau (equilibrated with the electronic kinetic temperature) in all cases in aluminum after deposition of 2 eV/atom absorbed dose. The difference at other deposited doses is comparably small, see Appendix. Thus, in the case of optical irradiation, the effect of electronic nonequilibrium on the electron-phonon coupling and atomic heating in aluminum appears to be relatively small. The same can be concluded for other studied metals: gold and tungsten (see Appendix).

Since in bulk metals electronic excitation leads to a phonon hardening effect [55], no nonthermal damage is observed in the current simulations. The situation may be different in nano-sized metallic samples where nonthermal phase transitions can take place in response to an increase of the electronic pressure and ensuing material expansion [55], but finite-sized samples are beyond the scope of the present work and may be a subject of future dedicated research.

Having studied the effect of the nonequilibrium in the wide range of possible relaxation times, let us now consider the experiment measuring the photon spectra emitted from L-shell decays in aluminum following irradiation with a 35-femtosecond free-electron laser (FEL) pulse of 92 eV photon energy (13.5 nm wavelength) [56]. In the soft X-ray regime, the photons are predominantly absorbed by the L-shell (2p atomic shell) of aluminum, with only a minor contribution of the conduction band photoabsorption. The L-shell holes formed during photoabsorption subsequently undergo decay, which is primarily Auger decay, with radiative decay contributing to a lesser extent. The radiative decays produce photons with energies defined by the ionization potential of the L-shell (~72 eV), which are detected experimentally. The detected spectrum is naturally averaged over the lifetime of the L-shell holes, which is ~40 fs in solid aluminum [56].

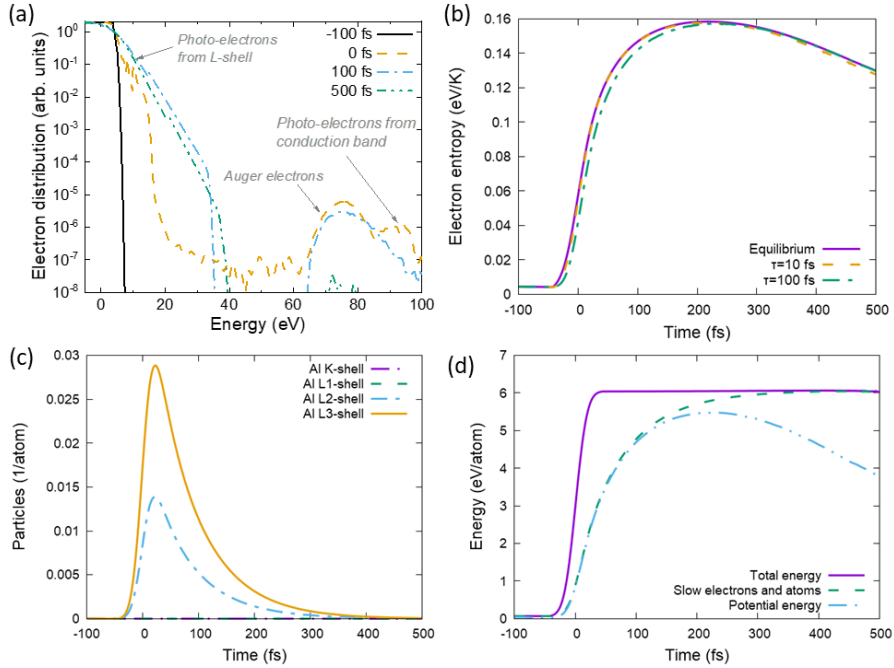

*Figure 4. Parameters of aluminum irradiated with 92 eV photon energy, 35 fs FWHM pulse, 6 eV/atom absorbed dose. (a) Electron distribution function. (b) Electronic entropy, simulated with various characteristic relaxation times of low-energy electrons. (c) The density of holes in various atomic shells. (d) Energies in various systems: total energy in the simulation box, energy of atoms and low-energy electrons, and potential energy of atoms (without the kinetic energy).*

We conducted simulations of aluminum irradiated with an FEL pulse with the parameters corresponding to those of the experiment and 500 atoms in the simulation box. Figure 4a illustrates the calculated electronic distribution function. The characteristic shape of the bump-on-hot-tail distribution, typical for FEL-irradiated materials, emerges during and after the FEL pulse [25,57]. The high-energy tail originates from the conduction-band photoabsorption during the pulse and is subsequently sustained by the Auger-decays of L-shell holes (see a detailed discussion in [25]). This tail lives as long as the L-shell holes are still present in the material, which may last for a few hundred femtoseconds (exponentially decaying with the characteristic time of ~ 40 fs, see Figure 4c)[58]. However, the low-energy fraction of electrons around the Fermi level is much closer to partial equilibrium, as was also discussed in [25]. These electrons participate in the radiative decays of L-shell holes, forming the experimentally observed spectrum.

The deviation of the electronic ensemble from equilibrium is not large, as seen by the electronic entropy (Figure 4b), even for the relaxation time of the low-energy electrons $\tau_{e-e}$=100 fs, however, the minority of high-energy electrons transiently stores some energy; a larger amount of energy is transiently stored in L-shell holes. Figure 4d shows that the energy of L-shell holes and high-energy electrons (the difference between the total energy plotted and the energy of atoms and slow electrons) transiently may be dominant. This figure also demonstrates that the potential energy of atoms starts to decrease due to heating, and the corresponding energy transfers to the kinetic energy of atoms, heating the system. It also can be seen that the total energy in the simulation box is conserved (apart from the deposition by the FEL pulse, centered around zero), validating the developed simulation scheme.

The calculated distribution function averaged over time with the transient density of L-shell holes (Figure 4), is then compared with the distribution function extracted from the experimental spectra (by division of the spectrum by the density of states). The results of this comparison are depicted in Figure 5, for the few relaxation times employed in XTANT-3 simulation. It is apparent from the figure that the closest match between the two distribution functions occurs for the characteristic times of ≤10 fs, setting the upper limit for the relaxation time to be used in simulations of aluminum (at least at comparable irradiation conditions).

The experimental peak at ~10 eV in Figure 5a is from impurities [59], and thus does not appear in the simulation. The minor deviation at energies below the Fermi level in Figure 5b seems to stem from the particular shape of the calculated density of states of the material, not the transient distribution function, which is the main subject of investigation here. The thin spikes at the lowest energies are experimental noise.

We also note that the electron relaxation times in other metals may be indirectly measured *via* the evolution of the optical coefficients, which are also sensitive to the particular shape of the electron distribution function. Optical or XUV measurements in irradiated samples thus allow for estimation of the average electron thermalization times [60,61].

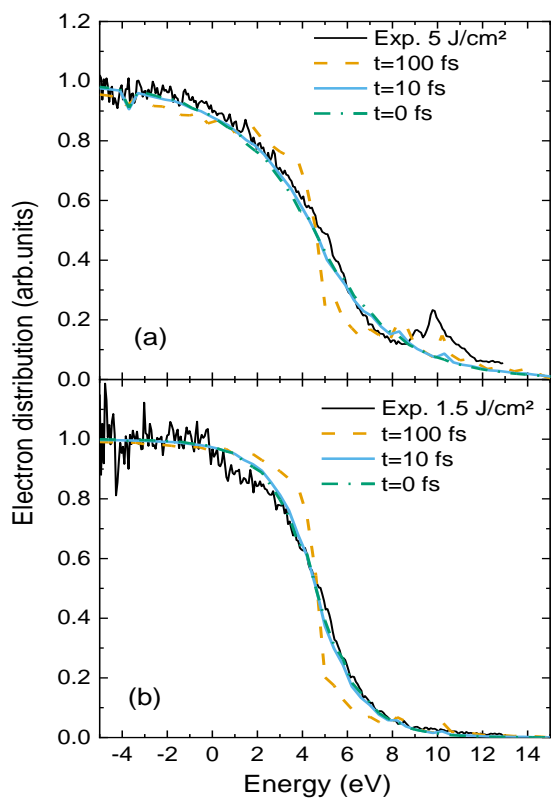

*Figure 5. Low-energy parts of the electron distribution in aluminum irradiated with 35 fs FWHM 92-eV photon energy laser pulse, simulated with XTANT-3 with various characteristic relaxation times, for the deposited doses of 1.8 eV/atom (a) and 5.8 eV/atom (b), compared with the experimental distributions extracted from the corresponding spectra from Ref. [56].*

Keep in mind that this is the characteristic relaxation time of only the low-energy fraction of electrons, whereas the *total* thermalization of the electronic system requires considerably more time (cf. Figure 4). The lifetime of the long nonequilibrium tail is defined by the slowest process, which in the case of excitation of the L-shell of aluminum is the Auger-decay time. This process continuously supplies the distribution function with new out-of-equilibrium electrons at the energies of ~72 eV for several hundred femtoseconds, as was previously investigated in [25] and experimentally confirmed in [58].

This comparison allows us to conclude that the relaxation time approximation, combined with MC simulation for high-energy electrons, is a reliable model, capable of describing experimental data with reasonable accuracy.

### 2. Semiconductors

We use silicon as the main target to study electronic nonequilibrium effects in semiconductors (compared with germanium). Here, again, the laser pulse of 10 fs FWHM with the photon energy of 2 eV is used. The deposited dose is set to 0.9 eV/atom, which is close to the nonthermal melting threshold in silicon [62]. Similar to the case of aluminum above, the distribution function of electrons in silicon forms a nonequilibrium shape during the laser pulse and secondary electron cascades, see Figure 6. The bandgap, seen by the positions of the points in the plotted distribution function, closes after a few hundred femtoseconds (for details, see e.g. Ref. [62]).

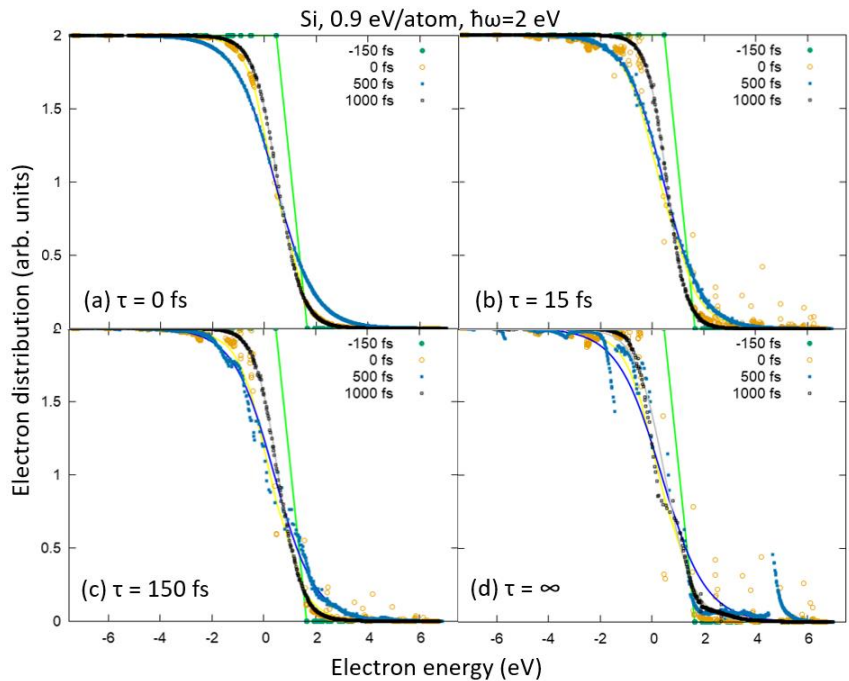

*Figure 6. Electronic distribution function in silicon irradiated with a pulse of 0.9 eV/atom, 2 eV photon energy, 10 fs FWHM. Simulations performed with different characteristic electronic relaxation times: (a) $\tau_{e\text{-}e}$=0 fs (instantaneous relaxation); (b) $\tau_{e\text{-}e}$=15 fs; (c) $\tau_{e\text{-}e}$=150 fs; (d) $\tau_{e\text{-}e} \rightarrow \infty$ (no electron-electron relaxation; Ehrenfest-like approximation). Symbols are the transient distribution functions, lines are the equivalent Fermi distributions.*

The thermalization takes longer for larger electron-electron relaxation times, as seen by the behavior of the electron entropy in Figure 7. Due to the presence of the bandgap at the beginning, before silicon nonthermally melts and turns metallic with the collapse of the gap, some energy is stored in electrons as the potential energy, which slows down electron thermalization. By the time of 1 ps, the thermalization of the electronic ensemble is almost completed even in the simulation with no electron-electron relaxation included ($\tau_{e-e} \to \infty$) – the same effect as described above for aluminum irradiation. This happens because, at the considered dose, ultrafast nonthermal melting of silicon takes place, leading to bandgap collapse and metallization of silicon [62]. Al lower doses, the limit of absent electron-electron thermalization ($\tau_{e-e} \to \infty$) leads to only partial thermalizations within each band, but not between the bands, as will be discussed in more detail below for the case of insulators (section III.3), and the case of germanium in the Appendix.

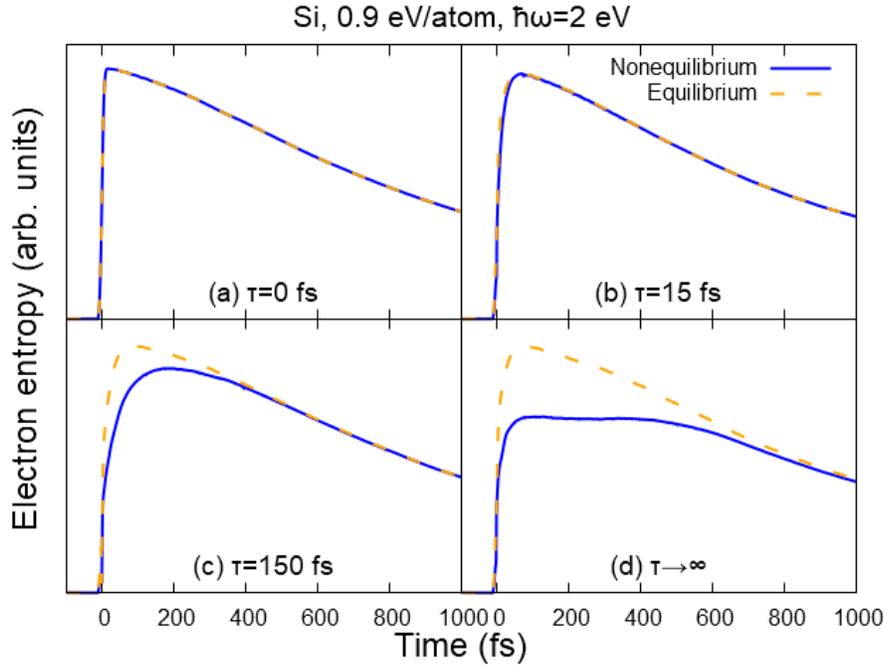

*Figure 7. Evolution of the electronic entropy in silicon irradiated with a pulse of 0.9 eV/atom, 2 eV photon energy, 10 fs FWHM. Simulations performed with different characteristic electronic relaxation times: (a) $\tau_{e-e}=0$ fs (instantaneous relaxation); (b) $\tau_{e-e}=15$ fs; (c) $\tau_{e-e}=150$ fs; (d) $\tau_{e-e} \to \infty$ (no electron-electron relaxation; Ehrenfest-like approximation). Solid lines are the entropies corresponding to the transient distribution functions, dashed lines are the entropy associated with the equivalent Fermi distributions (maximal entropy).*

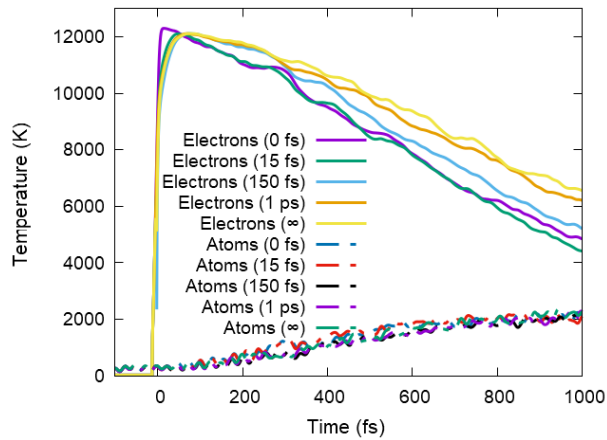

*Figure 8. Electronic and atomic kinetic temperatures in silicon irradiated with a pulse of 0.9 eV/atom, 2 eV photon energy, 10 fs FWHM. Simulations performed with different characteristic electronic relaxation times: $\tau_{e-e}=0$ fs, 15 fs; 150 fs, 1 ps, and $\tau \to \infty$ (no electron-electron*

Nonequilibrium in the electronic system affects the electron-phonon coupling in silicon stronger than in metals, see Figure 8: the electronic kinetic temperatures for large electron-electron relaxation times show a noticeable difference from the short-relaxation-time simulations. However, the atomic temperatures differ only mildly.

Nonthermal damage mechanisms are also affected by electronic non-equilibrium. Figure 9 shows that the nonthermal melting threshold first lowers with the increase of the electron-electron relaxation time, up to the times of $\tau_{e-e} \sim 150$ fs, where the damage threshold is ~0.85 eV/atom vs. ~0.9 eV/atom within the instantaneous thermalization ($\tau_{e-e} = 0$ fs) reported earlier [62]. For the thermalization times over $\tau_{e-e} \sim 250$ fs, the nonthermal damage threshold increases again, reaching ~0.9–1.0 eV/atom in the limit of $\tau_{e-e} \to \infty$.

In germanium, the situation is qualitatively similar, albeit the change in the damage threshold dose becomes more noticeable at larger thermalization times, see Figure 9. This could be expected, considering that Ge atoms are heavier than Si, and it takes longer for them to react to electronic excitation and melt the lattice nonthermally. So, longer-lasting electronic nonequilibrium is required to influence this process.

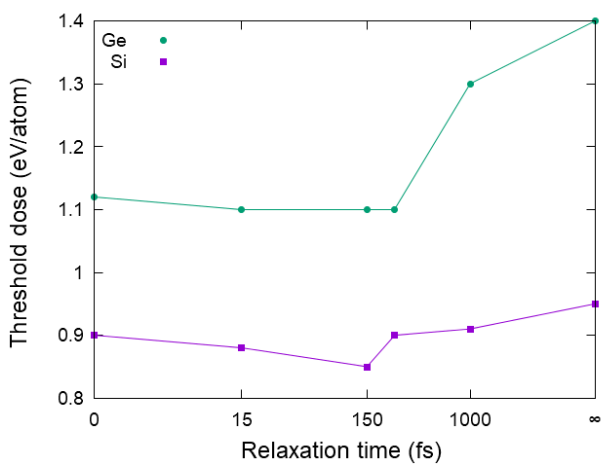

*Figure 9. Damage threshold dose in silicon and germanium irradiated with a pulse of 10 fs FWHM as a function of the characteristic electronic relaxation times.*

The nonlinear dependence of the damage threshold on the electron-electron thermalization time is mainly defined by the transient density of the electrons excited to the conduction band, see an example in Figure 10. Electrons, promoted to the conduction band by photoabsorption and relaxing *via* electron-electron scattering, excite secondary electrons into the conduction band, thereby increasing their density. When this channel is absent, fewer conduction band electrons are excited. With the increase of the relaxation time up to $\tau_{e-e} \sim 250$ fs, the density of the conduction band electrons holds over-the-threshold values (above ~4 % of the valence band electrons [62]) for a longer time, allowing the atoms to react and form nonthermal damage. However, the peak density simultaneously reduces, and for some values of the relaxation times ($\tau_{e-e} \sim 250$ fs) becomes insufficient to induce nonthermal melting (Figure 10). Thus, to reach the threshold electron density in simulations with larger relaxation times, a higher dose is required.

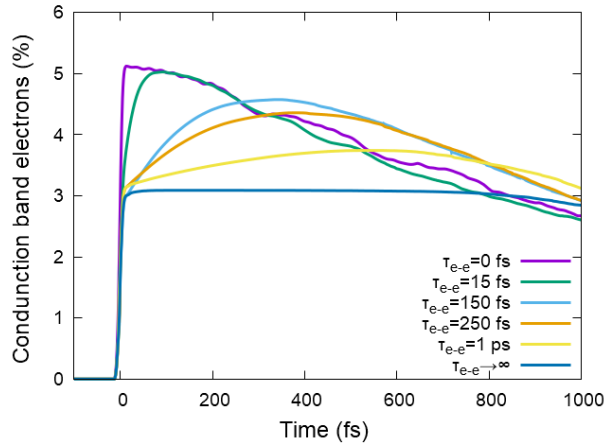

*Figure 10. The density of conduction band electrons (as a percentage of valence-band electrons) in silicon irradiated with a pulse of 0.9 eV/atom, 2 eV photon energy, 10 fs FWHM. Simulations performed with different characteristic electronic relaxation times: $\tau_{e-e}=0$ fs, 15 fs; 150 fs, 250 fs, 1 ps, and $\infty$ (no electron-electron relaxation; Ehrenfest-like approximation).*

It is interesting to note that the earlier results obtained with the instantaneous thermalization approximation ($\tau_{e-e} = 0$) predicted that the damage threshold in terms of the absorbed dose is independent of the photon energy (under the assumption of homogeneous photon absorption and no energy sinks from the sample) [62]. However, nonequilibrium simulation produces a different damage threshold that depends on the particular shape of the transient electron distribution function. Therefore, the threshold absorbed dose depends on the photon energy. We demonstrate it in the example of the largest effect of nonequilibrium, the limit of $\tau_{e-e} \rightarrow \infty$, see Figure 11.

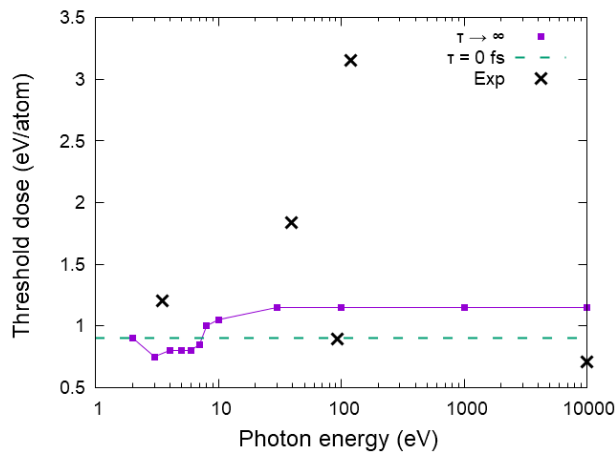

*Figure 11. Nonthermal melting threshold deposited dose in silicon irradiated with a pulse of 10 fs FWHM as a function of the photon energy. All simulations are performed with the characteristic electronic relaxation time $\tau_{e-e} \rightarrow \infty$ (no electron-electron relaxation; Ehrenfest-like approximation). The dashed horizontal line is the threshold dose in simulations with instantaneous electron thermalization ($\tau_{e-e}=0$ fs). The crosses are experimental data from [63] and references therein for XUV and x-rays, and from [64] for the optical pulse.*

For photon energies in the optical and ultraviolet range, there is a nonlinear dependence of the ultrafast nonthermal damage threshold. So, each particular photon energy requires a dedicated study. In contrast, for photon energies above the width of the valence band (~ 15 eV), the secondary cascades of high-energy electrons (photo-, Auger-, and impact-ionized) only differ in duration (see Ref. [65]), but the distribution of the secondary excited electrons does not differ significantly. Thus, with the increase of the photon energy above some ~15 eV, the damage threshold dose stays nearly constant up to hard X-rays, the highest photon energy of 10 keV studied here.

This result suggests that an indirect validation of the electron-electron relaxation time is potentially possible by experiments measuring the damage threshold. If all the other uncertainties of the simulation and experimental procedure were eliminated, the damage threshold dose would enable us to extract average electron-electron thermalization time by comparison with simulations. Unfortunately, the existing experimental data do not allow for discrimination between different characteristic thermalization times in the present calculations (see Figure 11). In Figure 11, the comparisons are shown for XUV and X-ray experiments with only one point for the optical pulse, which was converted from the excited electron densities from Ref. [64], since in most of the optical-laser experiments, the threshold is reported only in terms of the incident fluence. The nonlinear photoabsorption and other ensuing effects in the optical case make it difficult to convert the thresholds from incoming fluence to the absorbed dose [66].

### 3. Insulators

To study the effects of electron-electron equilibration in an insulator, we use $Al_2O_3$ (and diamond, in Appendix). Since, in the current implementation, XTANT-3 only includes linear photoabsorption, we cannot model photon energies smaller than the bandgap of an insulator, which is $E_{gap}$~9 eV in $Al_2O_3$. Thus, we chose an XUV pulse with a photon energy of 30 eV (wavelength of 41 nm). In this case, the high-energy electrons created due to photoabsorption, relax to the energies below the cutoff of 10 eV within ~1 fs, joining the low-energy fraction of electrons (utilizing Eqs. (4)) already during the laser pulse [65].

The evolution of the electronic distribution in irradiated $Al_2O_3$ is shown in Figure 12. Both, the transient nonequilibrium distribution and the equivalent equilibrium distribution function plotted have initially a large discontinuity at the bandgap. As the material starts to respond to the deposited energy, some energy levels are shifting into the gap, see Figure 13, and the electronic populations are changing [67]. In the case of fast thermalization ($\tau_{e-e} = 0$ fs and $\tau_{e-e} = 15$ fs), the kinks and spikes in the distribution function relax fast towards the Femi distribution, as expected. Slower thermalization time of $\tau_{e-e} = 150$ fs demonstrates persisting nonequilibrium at times of 1 ps.

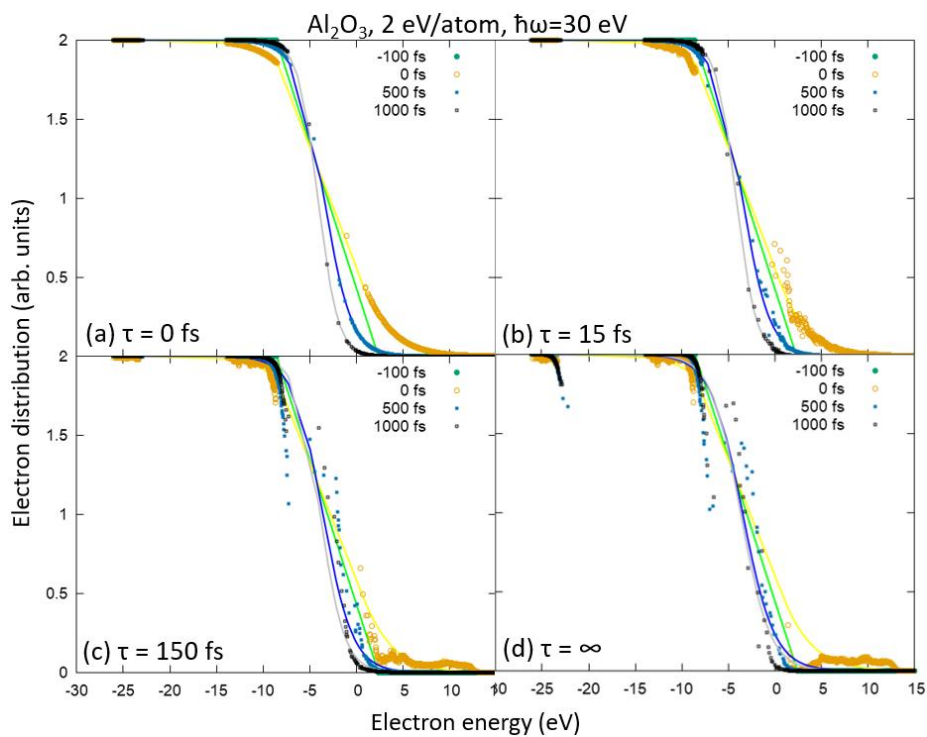

*Figure 12. Electronic distribution function in Al₂O₃ irradiated with a pulse of 2 eV/atom, 30 eV photon energy, 10 fs FWHM. Simulations performed with different characteristic electronic relaxation times: (a) τ_{e-e}=0 fs (instantaneous relaxation); (b) τ_{e-e}=15 fs; (c) τ_{e-e}=150 fs; (d) τ_{e-e}→∞ (no electron-electron relaxation; Ehrenfest-like approximation). Symbols are the transient distribution functions, lines are the equivalent Fermi distributions.*

In the absence of electron-electron scattering ($\tau_{e-e} \to \infty$), the electron interaction with phonons leads to only partial thermalization in the electronic ensemble, in contrast to the case of metals and highly-excited semiconductors discussed above. In an insulator, phonons cannot provide or accept sufficient energy for electrons to cross the bandgap, thus only separate intraband electron thermalizations are observed. Each band – conduction, valence, and subvalence – establishes its own separate distribution, close to the Fermi function (see Figure 12d). There is only a slow small leak of electrons from the conduction band to the valence band due to the lowering of the bands and occasional crossing of the valence and the conduction levels in the highly excited material, see Figure 13. Crossing or avoided crossing of energy levels allows for nonadiabatic electron transitions between them [68,69].

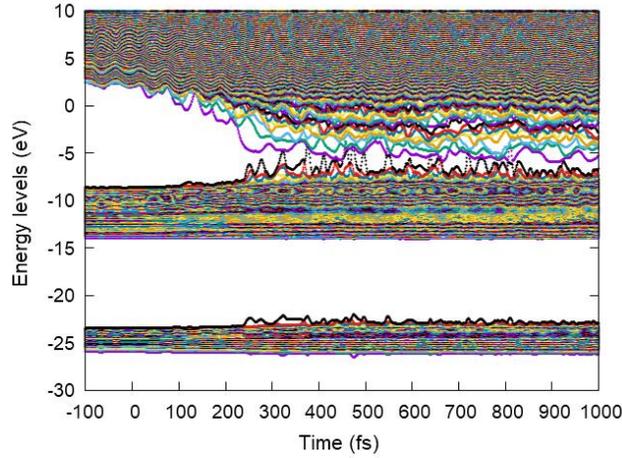

*Figure 13. Evolution of the electronic energy levels in Al₂O₃ irradiated with a pulse of 2 eV/atom, 30 eV photon energy, 10 fs FWHM, modeled with $\tau_{e-e} \rightarrow \infty$ (no electron-electron relaxation; Ehrenfest-like approximation).*

These processes are reflected in the electronic entropy being noticeably below the maximal value corresponding to the complete thermalization of all electrons, see Figure 14. The full electron thermalization within the Ehrenfest-like approximation ($\tau_{e-e} \rightarrow \infty$) does not take place in insulators at considered doses (the same effect manifests itself in semiconductors at doses below the bandgap collapse, see an example in Appendix). This illustration demonstrates that accounting for electron-electron thermalization is very important: models that neglect this effect may produce qualitatively different results. Figure 14 shows that the electron-electron thermalization may be captured in the relaxation time approximation with finite $\tau_{e-e}$, as implemented in XTANT-3.

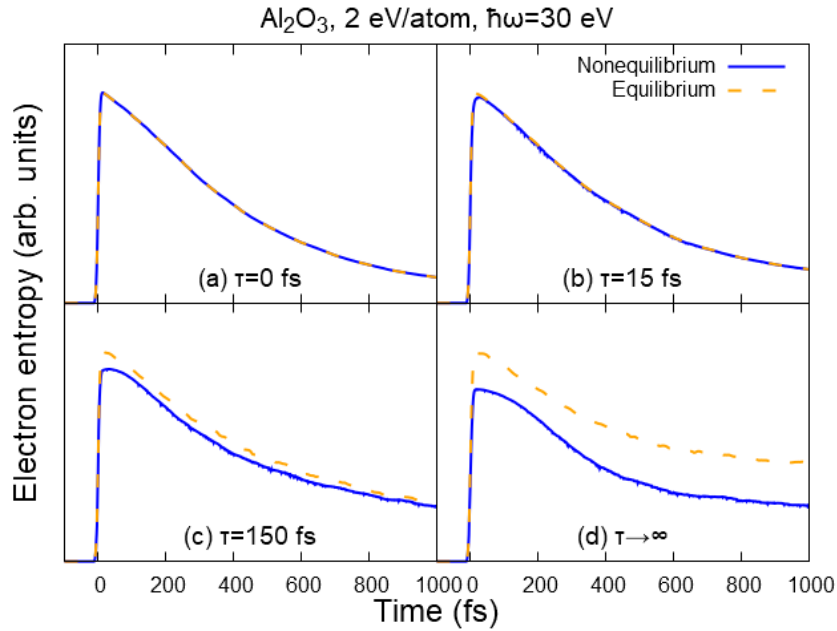

*Figure 14. Evolution of the electronic entropy in Al₂O₃ irradiated with a pulse of 2 eV/atom, 30 eV photon energy, 10 fs FWHM. Simulations performed with different characteristic electronic relaxation times: (a) $\tau_{e-e}=0$ fs (instantaneous relaxation); (b) $\tau_{e-e}=15$ fs; (c) $\tau_{e-e}=150$ fs; (d) $\tau_{e-e} \rightarrow \infty$ (no electron-electron relaxation; Ehrenfest-like approximation). Solid lines are the entropies corresponding to the transient distribution functions, dashed lines are the entropy associated with the equivalent Fermi distributions (maximal entropy).*

In the case without electron-electron relaxation ($\tau_{e-e} \to \infty$), a large amount of energy is stored as the potential energy of electrons: the energy that could be released if electrons had a mechanism to relax across the band gaps. The effect of it is twofold. On the one hand, the potential energy stored cannot be transferred to atoms/phonons *via* the nonadiabatic electron-ion coupling. This leads to slower heating of atoms and a lower final atomic temperature, see Figure 15. In the Ehrenfest-like approximation ($\tau_{e-e} \to \infty$), the kinetic electronic and atomic temperatures reached saturation by the time of ~1 ps and did not equilibrate, showing a failure of this approximation and the importance of finite electron relaxation time.

On the other hand, without direct electron-electron relaxation, valence holes and conduction band electrons last longer than in the simulation where electron-electron relaxation is included. That means, the electronic population on the bonding levels in the valence band is reduced, whereas the population of electrons on the antibonding states in the conduction band is increased, as can be seen by the persisting nonequilibrium shape of the distribution, recall Figure 12. The interatomic potential is directly affected by the electronic distribution on the levels (see Eq.(5)) [18,67]. This nonequilibrium electron distribution leads to ultrafast nonthermal phase transition in $Al_2O_3$ at doses lower than the equilibrium one: ~2.4 eV/atom in the case of the instantaneous thermalization ($\tau_{e-e} = 0$ fs) vs. ~2.2 eV/atom without electron-electron thermalization ($\tau_{e-e} \to \infty$). The state after the phase transition seems to be the same independently of the electron-electron thermalization time – transiently superionic state with the liquid oxygen subsystem and the solid aluminum lattice, for details see Refs. [18,67].

Thus, it can be concluded that the electronic nonequilibrium in insulators plays an important role in influencing both, nonadiabatic (electron-phonon) coupling and nonthermal phase transitions – an effect that the Boltzmann equation is capable of treating even within the relaxation-time approximation. The same effects are expected to take place in other covalent insulators and semiconductors where the nonthermal bandgap collapse may be induced by ultrafast irradiation. The ionic crystals may behave differently since they typically do not experience bandgap collapse under irradiation [70]. This topic is beyond the scope of the present work and may be studied in the future.

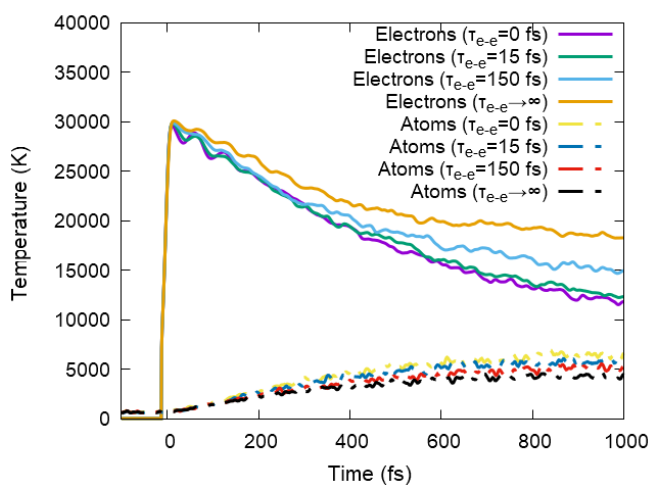

Figure 15. Electronic and atomic kinetic temperatures in $Al_2O_3$ irradiated with a pulse of 2 eV/atom, 30 eV photon energy, 10 fs FWHM. Simulations performed with different characteristic electronic relaxation times: $\tau_{e-e}=0$ fs, 15 fs, 150 fs, and $\infty$ (no electron-electron relaxation;

*Ehrenfest-like approximation). Solid lines are the electronic kinetic (equivalent) temperatures, and dashed lines are the atomic kinetic temperatures.*

## IV.  Conclusions

Materials response to ultrafast laser irradiation was studied theoretically using a comprehensive hybrid simulation tool XTANT-3. The evolution of the nonequilibrium electronic distribution function on the transient energy levels was included, which enabled accounting for the interplay of nonequilibrium, nonadiabatic, and nonthermal effects. The solution of the Boltzmann collision integrals developed is particle and energy conserving by construction. The model was applied to various classes of solids: metals, semiconductors, and insulators.

The results suggest that models based on Ehrenfest-like dynamics may be applied to describe the atomic response to irradiation of metals and semiconductors at doses above nonthermal melting, but they fail to describe insulators since electronic thermalization across the bands is excluded. Such models exhibit equilibration within the electronic ensemble mediated by the electron-phonon coupling but on much longer timescales than the actual thermalization *via* electron-electron interaction, which may be captured in the relaxation time approximation discussed here. The non-equilibrium distribution of electrons influences both, nonadiabatic electron-phonon coupling and nonthermal modification of the interatomic potential.

The nonequilibrium electron distribution leads to smaller electron-phonon coupling and, correspondingly, slower atomic heating in all studied cases (all materials, irradiation doses, and photon energies). In metals, the difference in the atomic response simulated with different models is relatively small: only slightly slower atomic heating is observed for larger electron-electron relaxation times. Comparison of the simulated electron distribution function with the experimental emission spectra enabled estimating the thermalization time in aluminum as ≤10 fs. However, longer-lasting non-equilibrium may be expected in some cases [71] or, perhaps, under long-pulse irradiation continuously driving the electronic system out of equilibrium.

In semiconductors, the effect of electronic nonequilibrium on the electron-phonon coupling is similar, but it also plays a role in nonthermal melting. Depending on the electron-electron relaxation time, the nonthermal damage threshold may be lowered or increased, since electron-electron scattering affects the excitation of electrons across the bandgap, and the conduction-band electrons and valence holes are the main driving force behind the nonthermal melting. It, again, demonstrates the importance of accounting for the finite time of the electron-electron thermalization in reliable simulations.

In insulators, electronic nonequilibrium leads to a lower nonthermal damage threshold, since unrelaxed electrons excited to the conduction band affect the interatomic potential stronger than that in the case of assumed instantaneous electronic thermalization. Without electron-electron thermalization, there is no mechanism for electrons to hop between the conduction and the valence band, which strongly affects both: slows down electron-phonon coupling precluding electron-ion thermalization, and changes the threshold of nonthermal phase transitions. Thus, an appropriate model of the insulator's response to ultrafast irradiation should include all three effects: electronic nonequilibrium, nonadiabatic electron-ion coupling, and nonthermal evolution of interatomic potential.

## Data availability statement

The code XTANT-3 used in this work is publicly available in [29] or from https://github.com/N-Medvedev/XTANT-3.

## Acknowledgments

The author thanks V. Lipp, R. Santra, O. Vendrell, A.E. Volkov, R. Voronkov, and B. Ziaja for helpful discussions. The financial support from the Czech Ministry of Education, Youth and Sports (grants No. LTT17015, LM2023068, and No. EF16_013/0001552) is gratefully acknowledged. Computational resources were supplied by the project "e-Infrastruktura CZ" (e-INFRA LM2018140) provided within the program Projects of Large Research, Development and Innovations Infrastructures.

## Conflict of interest

The author declares no conflict of interest, financial or otherwise.

## Appendix

Figure 16 shows the evolution of the electronic and atomic kinetic temperatures in aluminum irradiated with 10 fs FWHM laser pulse, 2 eV photon energy, and various absorbed doses. Two simulations are compared: with the electron-electron relaxation time of $\tau_{e-e} = 15$ fs vs. $\tau_{e-e} \to \infty$. For all deposited doses, the difference between the two simulation schemes is comparably small. Due to the dependence of the electron-ion (electron-phonon) coupling on both, electronic and atomic temperatures, nonlinear behavior of the temperatures is observed: for higher deposited doses (peak electronic kinetic temperatures), the temperatures equilibrate faster than for lower doses, see discussions in [24,72].

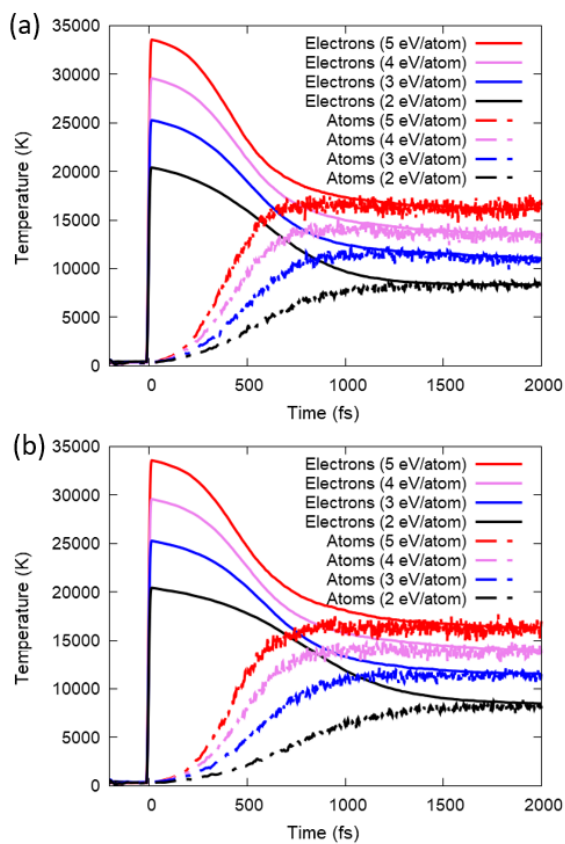

*Figure 16. Electronic and atomic temperatures in aluminum irradiated with a pulse of various deposited doses, 2 eV photon energy, 10 fs FWHM. Simulations performed with different characteristic electronic relaxation times: (a) $\tau_{e-e}=15$ fs; (b) $\tau_{e-e}\rightarrow\infty$ (no electron-electron relaxation; Ehrenfest-like approximation). Solid lines are the electronic kinetic (equivalent) temperatures, and dashed lines are the atomic kinetic temperatures.*

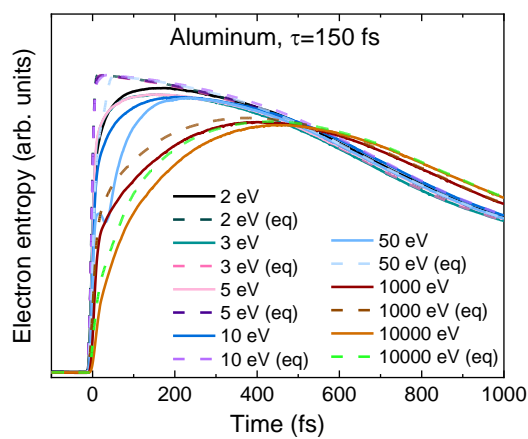

*Figure 17. Electronic entropy in aluminum irradiated with a pulse of various photon energy, 10 fs FWHM, 2 eV/atom absorbed dose. Simulations performed with the characteristic electronic relaxation time of $\tau_{e-e}=150$ fs. Solid lines are the nonequilibrium electronic entropy, and dashed lines are the maximum (equilibrium) counterparts (marked with "eq").*

The photon energy of the irradiation pulse also affects the electron behavior, as can be seen in Figure 17. It shows that with an increase of the photon energy from the optical to extreme ultraviolet range (above ~10 eV), the equilibrium (maximal) electronic entropy peak shifts to longer times due to electron cascades. The nonequilibrium entropy increase in all cases takes longer, as defined by the finite relaxation time (in this case 150 fs). As photon energy increases above the L-shell ionization potential (~72 eV in aluminum), the electron cascades become significantly longer due to the Auger decays, as discussed in the main text, section III.1. Further increase of the photon energy and, correspondingly, an increase of the photo-electron energy, prolongs the cascades.

Figure 18 presents the kinetic electronic and atomic temperatures in gold and tungsten irradiated with 10 fs FWHM laser pulse, 2 eV photon energy, and 2 eV/atom absorbed dose, using various electron-electron relaxation times. The difference among all the cases is relatively small, indicating that the electronic nonequilibrium does not influence strongly the electron-ion coupling in these metals, supporting the conclusions in the main text. It appears to be the smallest in the case of the slowest electron-phonon coupling (typically, heavier elements [14]). It could be expected, considering the ratio of the electron relaxation times and the characteristic timescales of the L-electron-ion thermalization in the studied materials.

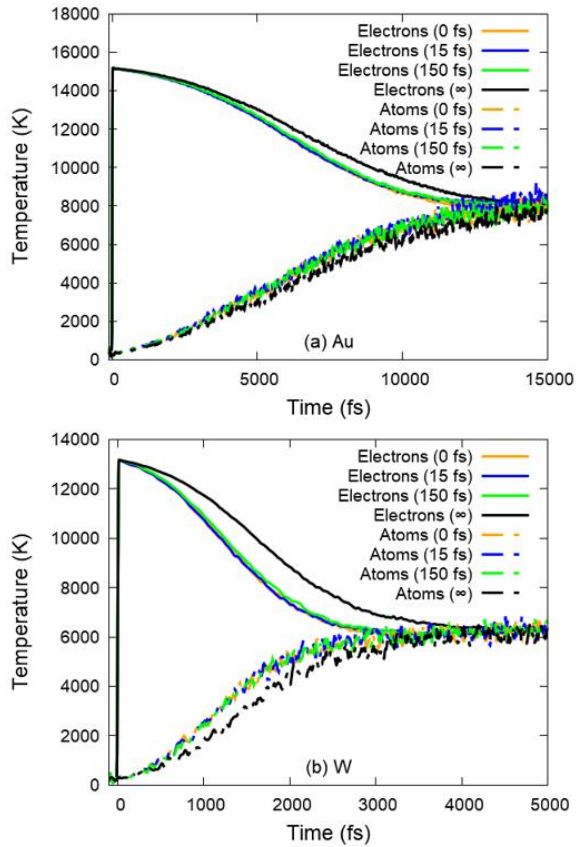

*Figure 18. Electronic and atomic temperatures in (a) gold and (b) tungsten irradiated with a pulse of various deposited doses, 2 eV photon energy, 10 fs FWHM, and different characteristic electronic relaxation times. Solid lines are the electronic kinetic (equivalent) temperatures, and dashed lines are the atomic kinetic temperatures.*

The electronic thermalization in semiconducting germanium can be seen in the evolution of the electron entropy in Figure 19. For this case, irradiation with a 1.2 eV/atom deposited dose is used, which is above the nonthermal damage threshold for short electron thermalization times, but below it for long thermalization times (see Figure 9). Thus, in cases of the thermalization times of 0 fs, 15 fs. And 150 fs, the bandgap collapses, turning irradiated germanium metallic, whereas for $\tau_{e-e} \to \infty$, there is no bandgap collapse (at the studied timescales of up to 2 ps). This effect results in the observable electronic thermalization for the finite shown relaxation times (Figure 19a,b, and c – the transient entropy coincides with the maximal, thermalized one, after some time), whereas the absent electron-electron thermalization, $\tau_{e-e} \to \infty$, with the present bandgap prevents electronic thermalization (Figure 19d, transient entropy does not reach the maximal one). The same effect was discussed in the case of insulators in section III.3. Thus, we conclude that the electron-electron relaxation is also necessary to include in a reliable model of semiconductors, especially in the case of low-dose irradiation, below the dose leading to the bandgap collapse. Ehrenfest-like approximations (and the BO approximation by extension), excluding electron-electron thermalization, may lead to qualitative discrepancies.

Figure 20 shows electron entropy in an insulating diamond irradiated with 2 eV/atom absorbed dose, 30 eV photon energy, similar to the case of $Al_2O_3$ (cf. Figure 14). Diamond undergoes ultrafast nonthermal phase transition within ~150-200 fs into a semimetallic state with the bandgap collapse [10]. Thus, the electronic thermalization proceeds similarly to the case of semiconducting silicon – the collapsed bandgap leads to electronic thermalization mediated by the electron-phonon coupling, which leads to equilibration of the electronic and ionic (phononic) temperatures. Although a very strong coupling in diamond leads to fast electron thermalization (in the case of above-threshold dose deposition), still the case of $\tau_{e-e} \to \infty$ produces slower thermalization than finite electron-relaxation times. In the case of below-threshold deposited doses (not shown), the interband relaxation does not take place in the case of $\tau_{e-e} \to \infty$, similar to the discussed $Al_2O_3$, and the finite electron relaxation time is necessary to include in the model (Figure 20a-c and Figure 14a-c).

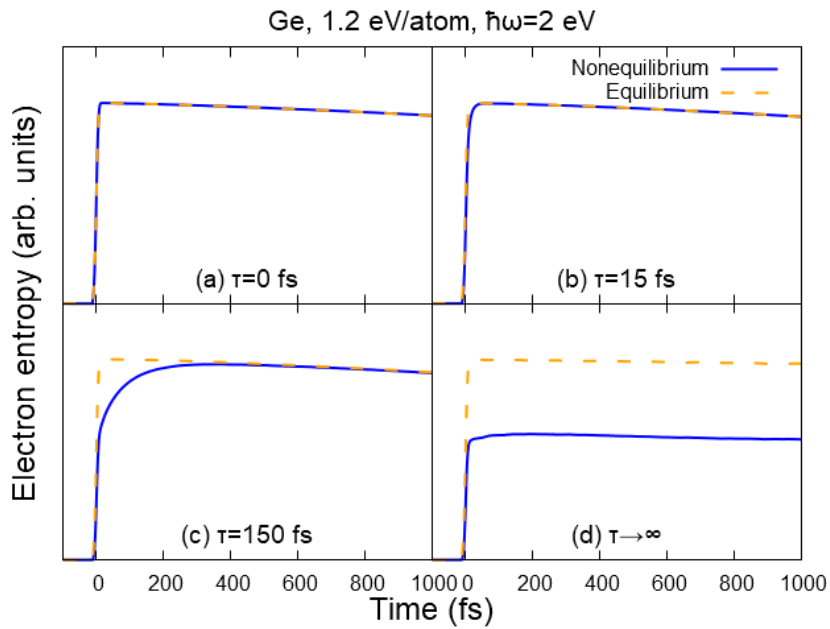

*Figure 19. Evolution of the electronic entropy in germanium irradiated with a pulse of 1.2 eV/atom, 2 eV photon energy, 10 fs FWHM. Simulations performed with different characteristic electronic relaxation times: (a) $\tau_{e-e}$=0 fs (instantaneous relaxation); (b) $\tau_{e-e}$=15 fs; (c) $\tau_{e-e}$=150 fs; (d) $\tau_{e-e} \to \infty$ (no electron-electron relaxation; Ehrenfest-like approximation). Solid lines are the entropies corresponding to the transient distribution functions, dashed lines are the entropy associated with the equivalent Fermi distributions (maximal entropy).*

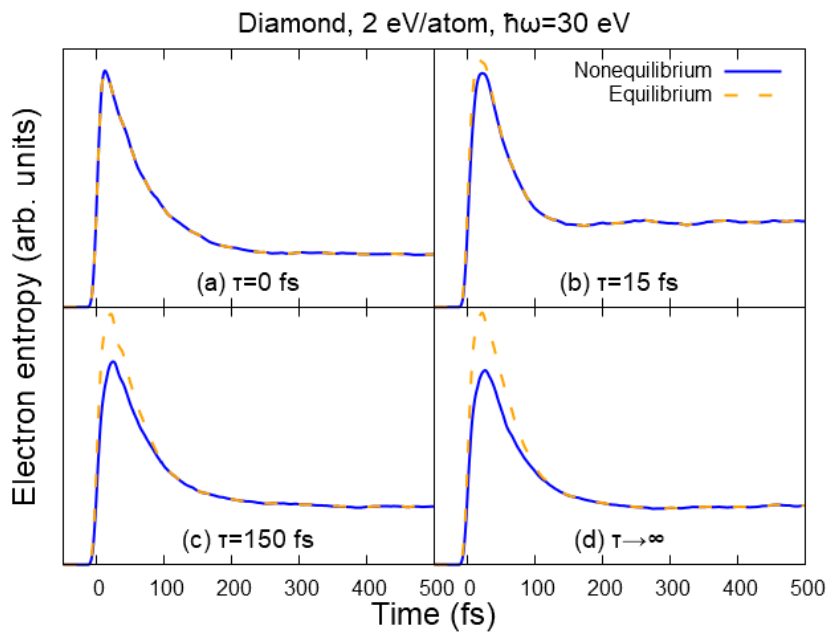

*Figure 20. Evolution of the electronic entropy in diamond irradiated with a pulse of 2 eV/atom, 30 eV photon energy, 10 fs FWHM. Simulations performed with different characteristic electronic relaxation times: (a) $\tau_{e-e}$=0 fs (instantaneous relaxation); (b) $\tau_{e-e}$=15 fs; (c) $\tau_{e-e}$=150 fs; (d) $\tau_{e-e} \rightarrow \infty$ (no electron-electron relaxation; Ehrenfest-like approximation). Solid lines are the entropies corresponding to the transient distribution functions, dashed lines are the entropy associated with the equivalent Fermi distributions (maximal entropy).*